\documentclass{kluwer}
\usepackage{epsfig}

\newcommand{\citeN}[1]{\citeauthor{#1}\ (\citeyear{#1})}
\newcommand{\citeNP}[1]{\citeauthor{#1},\ \citeyear{#1}}

\newcommand{\apj}{{\it Astrophys.~J.}}

\newcommand{\aap}{{\it Astron.~Astrophys.}}

\newcommand{\an}{{\it Astr.~Nachr.}}

\newcommand{\JFM}{{\it J.~Fluid~Mech.}}		
\newcommand{\JGR}{{\it J.~Gephys.~Res.}}

\newcommand{\mnras}{{\it Monthly Notices Roy.\ Astron.\ Soc.}}

\newcommand{\solphys}{{\it Solar~Phys.}}

\newcommand{\ujlisti}{
\itemsep=0 em
\parsep=0.5 em
\partopsep=0.25 em
\topsep=0 em}
\newcommand{\ujlistii}{
\itemsep=0 em
\parsep=0.5 em
\partopsep=0.25 em
\topsep=0 cm}

\newcommand{\<}{\begin{equation} }
\renewcommand{\>}{\end{equation} }

\def\epsi{\epsilon}

\newcommand{\pdv}[2]{\frac{\partial #1}{\partial #2}}

\def\ga{\mathrel{\mathchoice {\vcenter{\offinterlineskip\halign{\hfil
 $\displaystyle##$\hfil\cr>\cr\sim\cr}}}
 {\vcenter{\offinterlineskip\halign{\hfil$\textstyle##$\hfil\cr
 >\cr\sim\cr}}}
 {\vcenter{\offinterlineskip\halign{\hfil$\scriptstyle##$\hfil\cr
 >\cr\sim\cr}}}
 {\vcenter{\offinterlineskip\halign{\hfil$\scriptscriptstyle##$\hfil\cr
 >\cr\sim\cr}}}}}

\newcommand{\omegaBV}{N_{\mbox{\scriptsize BV}}}
\newcommand{\rbcz}{r_{\mbox{\scriptsize bcz}}}

\newcommand{\nucr}{\nu_{{\mbox{\scriptsize cr}}}}
\newcommand{\numax}{\nu_{{\mbox{\scriptsize max}}}}

\newcommand{\omegacyc}{\omega_{\mbox{\scriptsize cyc}}}
\newcommand{\Hskin}{H_{\mbox{\scriptsize skin}}}

\begin{document}                                                                                   
\begin{article}
\begin{opening}

\title{A CONSISTENT ONE-DIMENSIONAL MODEL FOR THE TURBULENT TACHOCLINE}

\author{K. \surname{Petrovay}}
\institute{E\"otv\"os University, Dept.~of Astronomy, Budapest, Pf.~32,
	   H-1518 Hungary\\
	   and\\
	   Center for Turbulence Research, NASA Ames Research Center, Moffett
Field, California 94035} 

\date{[{\it Solar Physics} {\bf 215}, 17--30]}

\runningtitle{A Model for the Turbulent Tachocline}
\runningauthor{K. Petrovay}

\pubyear{2003}

\begin{abstract}

The first consistent model for the turbulent tachocline is presented, with the
turbulent diffusivity computed within the model instead of being specified
arbitrarily. For the origin of the 3D turbulence a new mechanism is proposed.
Owing to the strongly stable stratification, the mean radial shear is stable,
while the horizontal shear is expected to drive predominantly horizontal,
quasi-2D motions in thin slabs. Here I suggest that a major source of
3D overturning turbulent motions in the tachocline is the {\it secondary shear
instability} due to the strong, random vertical shear arising between the
uncorrelated horizontal flows in neighbouring slabs. A formula for the vertical
diffusivity due to this turbulence, equation~(\ref{eq:nucr}), is derived and
applied in a simplified 1D model of the tachocline. It is found that Maxwell
stresses due to an oscillatory poloidal magnetic field of a few hundred gauss
are able to confine the tachocline to a thickness below 5 Mm. The
integral scale of the 3D overturning turbulence is the buoyancy scale, on the
order of 10\,km and its velocity  amplitude is a few m/s, yielding a vertical
turbulent diffusivity on the order of $10^8\,$cm$^2/$s. 

\end{abstract}


\end{opening}

\section{Introduction}

In most current dynamo models (see e.g. \citeNP{Petrovay:SOLSPA}, for a
review), the strong toroidal magnetic field responsible for solar activity is
generated in the thin rotational shear layer below the convective zone (CZ),
known as the tachocline. To hold sufficient magnetic flux to explain the flux
observed at the surface without unrealistically high field strengths, this
layer should be at least a few megameters thick. On the other hand, a magnetic
field oscillating with a circular frequency $\omegacyc=2 \pi /P$, ($P= 22$
years is the solar cycle period) is known to penetrate a conductive medium only
down to a skin depth of 
\< \Hskin=(2\eta/\omegacyc)^{1/2}             \label {eq:skin} \>
where $\eta$ is the magnetic diffusivity. If $\Hskin > 1\,$Mm then we conclude
from this that for present-day dynamo models to work, we must have $\eta\ga
10^8\,$cm$^2/$s in the tachocline layer. For comparison, the molecular magnetic
diffusivity is $\sim 10^3\,$cm$^2/$s in this part of the Sun, while the
turbulent diffusivity in the convective zone is estimated to be
$10^{12}$--$10^{13}\,$cm$^2/$s.

Helioseismic evidence indicates that at low heliographic latitudes the middle
of the tachocline is situated at $0.69\,R_\odot$ ($1\,R_\odot =698\,$Mm is the
solar radius) and its thickness is less than a few percent of $R_\odot$
(\citeNP{Kosovichev:tachocline}). There are indications that at high latitudes
the tachocline is situated at slightly shallower depths, $r=0.705R_\odot$, and
it may also be marginally thicker (\citeNP{Basu+Antia:tachovar}). Comparing
this with the radius of the bottom of the adiabatically stratified convective
zone, $0.71\,R_\odot$, it is clear that most of the tachocline lies in the
radiative interior, especially near the equator, so the relatively high value
of the (presumably turbulent) magnetic diffusivity inferred above is puzzling.
Hence, there is clearly a need to understand the nature of small-scale
turbulent motions in the tachocline. Unfortunately, while the thermal
stratification in the tachocline is relatively well known, its fluid dynamical
properties, including the precise profile of the rotational flow $v(r,\theta)$,
the meridional flow, and the characteristics of turbulence, are very poorly
constrained by observations. 

The tachocline is thus basically an MHD shear flow in the azimuthal direction
in a thin, rotating spherical shell with both radial and latitudinal shear and
a strongly stable stratification (Richardson number $\sim 10^3$). Theoretical
understanding and experimental knowledge regarding turbulence in stratified
shear flows has been reviewed by \citeN{Hopfinger} and \citeN{Thorpe}. These
reviews ephasize the tendency towards two-dimensional, predominantly horizontal
motions in such situations, the vertical motions being limited to the small
{\it buoyancy scale} (see below).  In line with the concepts outlined in those
reviews, the present paper to relies on theoretical arguments concerning the
properties of turbulent motions in such conditions. On the basis of such
arguments, simplified one- or two-dimensional models may be constructed for the
mean flow, or appropriate subgrid closure schemes may be constructed for full
3D large-eddy simulations. While such arguments have been attempted in the past
(\citeNP{Spiegel+Zahn}, \citeNP{Gough+McIntyre}),  in the lack of detailed
turbulence calculations to be used for subgrid closures, all tachocline models
published to date have either simply ignored turbulence
(\citeNP{Rudiger+Kichat:thin.tacho}, \citeNP{McGregor+Char},
\citeNP{Garaud:tacho1}, \citeNP{Cally}), or assumed arbitrary fixed scalar
turbulent diffusivities in 2D mean flow models (\citeNP{Dajka+Petrovay:mgconf},
\citeyear{Dajka+Petrovay:fast1}, \citeNP{Dajka:fast2}) and in 3D LES
(\citeNP{Miesch:tacho1}, \citeyear{Miesch:tacho2}).

The aim of the present work is to attempt to remedy this situation by
considering, on the basis of the known stability criteria, the possible 
sources of turbulence in a strongly stably stratified shear flow with both
vertical and horizontal shear, and discussing the expected properties of the
turbulence generated by it, on the basis of a dimensional analysis of the
$K$--$\epsi$ equations. As our analysis does not consider the effects of
spherical geometry, rotation, meridional circulation, non-adiabatic convective
overshoot, or magnetic instabilities, it should only be regarded as a fist step
towards a more comprehensive theoretical analysis of the problem of turbulence
in the solar tachocline. These theoretical arguments can be found in Section~2.
Then, in Section~3, as an illustration of the use of such theoretical
considerations, our prescription for calculating the turbulent diffusivity in
the tachocline, equation (\ref{eq:nucr}), is incorporated in a simplified
one-dimensional model for the tachocline. The results show that Maxwell
stresses due to an oscillatory poloidal magnetic field of a few hundred gauss
(a rather moderate value) are able to confine the tachocline to a thickness
below 5 megameters. The conditions of validity of our treatment are discussed
in Section~4. Finally, Section~5 concludes the paper by comparing our suggested
picture of tachocline turbulence with numerical simulation results and
discussing the implications of the results.

\section{A possible source of turbulence: the secondary shear instability}

The thermal stratification of the Sun is quite accurately known from a
comparison of standard solar models with helioseismic inversion results.  The
radiative interior below the CZ is characterized by significant
negative values of $\Delta\nabla$. For rough estimates, a useful approximation
in the tachocline region (i.e. near the top of the radiative zone) is
$\Delta\nabla\sim -0.015\,z$[Mm], where $z=\rbcz-r$ is the depth below the
bottom of the convective zone, at a radius value of $\rbcz=0.71R_\odot$. On the
other hand, the pressure scale height in this region is $H_P=-P\,dz/dP\sim
50\,$Mm. With $g=5\cdot 10^4\,$cm$^2/$s, this yields a Brunt-V\"ais\"al\"a
frequency 
\< \omegaBV^2\,[\mbox{s}^{-2}]=-\Delta\nabla g/{H_P}\sim 1.5\cdot 10^{-7}\,
   z\,\mbox{[Mm]}, \>
i.e. at a depth of a few megameters $\omegaBV\sim 10^{-3}\,$s$^{-1}$. 

A displaced mass element will then oscillate around its equilibrium position
under the action of buoyancy on a timescale $\omegaBV^{-1}\sim 10^3\,$s. The
amplitude of the oscillation is clearly $\sim v_z/\omegaBV$, so in the presence
of turbulent motions, these motions will be limited to a vertical scale
$l_b=K^{1/2}/\omegaBV$, called the {\it buoyancy scale}. ($K=\overline{v_z^2}$
is the kinetic energy in the vertical component of motions.) On the other hand,
as an elementary estimate gives $\nu\sim K^2/\epsilon$ for the vertical
turbulent diffusivity, one has $l_b\sim (\nu/\omegaBV)^{1/2}$. Vertical
overturning motions on scales exceeding $l_b$ will be strongly damped by
gravity wave emission.

The pole-equator difference in the rotational rate of the convective zone is
about $30\,$\% of the equatorial rotation rate. Taking half of this value to be
the characteristic amplitude of the differential rotation (cf. eq.
(\ref{eq:vhdef}) below), we have a differential rotation amplitude $\Delta
v_0\sim 3\cdot10^4\,$cm/s at the top of the tachocline. This value is clearly
also roughly the amplitude of the vertical velocity difference across the
tachocline, so the characteristic vertical shear is 
\< S\sim\Delta v_0/H\sim 3\cdot 10^{-5}\,\mbox{s}^{-1}
\>
where $H$ is the scale height of $\Delta v$, as given by equation
(\ref{eq:vhdef}) below. Helioseismic calibrations
(\citeNP{Basu+Antia:tachovar}) indicate $H\sim 10\,$Mm. This yields a
Richardson number Ri\,$=\omegaBV^2/S^2\sim 10^3$. This enormous value shows
that the vertical shear cannot directly drive turbulence in the tachocline. 

Buoyancy, however, cannot stabilize the horizontal shear. While linear
stability analysis (\citeNP{Dziembowski+Kosovichev:difrot.stab}, 
\citeNP{Charbonneau+:tacho.HDstab}) shows that the horizontal shear is close to
marginal stability in the 2D nonmagnetic case,  nonlinear, 3D effects and
magnetic fields are known to lead to strong instability
(\citeNP{Gilman+Dikpati:unstable.tacho},
\citeNP{Dikpati+Gilman:shallow.alpha}).  The motions driven by the horizontal
shear instability are predominantly horizontal, and their spatial scale is 
\< l_h\sim R_\odot ,
\>
On the basis of general experience with shear flow turbulence\footnote{In fact
it is not obvious that this experience is a useful guide under conditions in
the solar tachocline. This point will be discussed further in Sect.~4 below.},
the velocity  scale $v_h$ of the horizontal motions driven by the shear
instability may be  assumed to be close to the amplitude $\Delta v$ of the
horizontal shear at the  given depth:
\< v_h\sim\Delta v={[\omega(z,\theta=0)-\omega(z,\theta=\pi/2)]}R_\odot/2
   \label{eq:vhdef}       \>
Similarly, their correlation time is assumed to be 
\< l_h/v_h\sim R_\odot/\Delta v  
   \label{eq:thdef}       \>

Overturning turbulent motions in the vertical direction are impeded by the
stable stratification, their scale being limited to $l_b$. Owing to the low
vertical diffusivity and the finite correlation time, the horizontal
motions will be characterized by a limited vertical correlation length 
\< l_c\sim(\nu l_h/v_h)^{1/2} . \label{eq:lc} \>
The random horizontal flows driven by the shear will then be limited to thin
sheets of thickness $l_c$, the motion in neighbouring sheets being independent.
This will give rise to random vertical shear between neighbouring sheets, of
amplitude 
\< S_2\sim v_h/l_c \sim  (v_h^3/\nu l_h)^{1/2}  \>
This {\it secondary vertical shear} is much stronger than the primary (mean)
vertical shear, the corresponding Richardson number being 
\[ \mbox{Ri}_2=g\Delta\nabla \nu l_h/H_P v_h^3 . \]
 Substituting here the characteristic values of the
parameters, we find that Ri$_2<0.25$, i.e. the secondary shear is unstable, if 
\< \nu < \nucr= \frac{v_h^3}{4 l_h\omegaBV^2} 
    \simeq 10^{-4} \frac{v_h^3 \mbox{[CGS]}}{z \mbox{[Mm]}}  
   \label{eq:nucr}  \>
In a depth of a few megameters this value is $\nucr\sim 10^8\,$cm$^2/$s, much
higher than the molecular value, so we expect that, in the absence of other
sources of turbulence, the secondary shear is strongly unstable.

What is the characteristic amplitude of the turbulent motions driven by the
secondary shear instability? In principle, this could be derived from a
$K$--$\epsilon$ model (or, more, generally, from a Reynolds stress model 
(\citeNP{Speziale}).
Assuming plane parallel geometry for simplicity, the relevant equations are of
the general form 
\< \pdv Kt=P_K-D_K-\pdv{F_K}z
\>
\< \pdv \epsi t=P_\epsi-D_\epsi-\pdv{F_\epsi}z
\>
Here, the non-local fluxes or third order moments are
\< F_K=\overline{v_z^3}
\>
\< F_\epsi=\overline{v_z\epsi_l}
\>
$\epsi_l$ being the local dissipation rate, while $\epsi$ is the mean
dissipation. The production terms are usually modelled as
\< P_K=\frac\nu 2 S_2^2  \>
\< P_\epsi=C_P \frac\epsi K P_K   \>
while, assuming an ideal gas and the Boussinesq approximation
$\rho'/\rho=-T'/T$ (prime denotes fluctuations), the  dissipation/sink terms
read
\< D_K=\epsi+g\overline{v_z T'}  \label{eq:Dk} \>
\< D_\epsi= C_{D1} \frac{\epsi^2}{K} +C_{D2} \frac\epsi K g\overline{v_z T'} \>
(The first terms on the r.h.s. represent viscous dissipation, while the second
terms correspond to gravity wave emission. Note that in a subadiabatic
environment $\overline{v_z T'}>0$, i.e. downmoving fluid parcels are hotter than
average.)

Performing a dimensional analysis on these equations we find that the diffusive
timescale, corresponding to the non-local terms, is $d^2/\nu\ga d^2/\nucr\sim
10^{10}\,$s, while the timescale associated with the shear production term is
$K/\nu S_2^2\sim\omegaBV/S_2^2\sim \omegaBV l_h\nu/v_h^3\sim 10^{-5}\nu <
10^3\,$s. This implies that the transport terms can be neglected in the
$K$-$\epsi$ equations. Under such conditions the equations have no stationary
solution, as the values of the constants $C_P, C_{D1}$ and $C_{D2}$ are
different in general. The intensity of turbulence will then keep increasing
until the turbulent diffusivity reaches the critical value $\nucr$, when
further shear production is switched off. 

We thus conclude that the secondary vertical shear instability can be expected
to drive overturning turbulence to the level $\nu=\nucr$ on a short timescale.
The turbulence generated by this mechanism may then be crudely represented by
the vertical diffusivity value given by eq. (\ref{eq:nucr}).

\section{Tachocline model}

\subsection{Model equations}
We now proceed to develop a simple one-dimensional model for the solar
tachocline, assuming that the secondary shear instability discussed in the
previous section is the only source of turbulence in the tachocline region. Our
computational domain will be restricted to the top of the radiative interior,
below $\rbcz$.

The convective zone is characterized by extremely high turbulent diffusivities.
Due to the Coriolis force, the Reynolds stress tensor also has significant
nondiagonal components in the convective envelope. These components imply an
angular momentum transport which is thought to be the main driver of solar
differential rotation. Based on the discussion of the previous section we
expect that the amplitude of turbulence in the tachocline is much lower than in
the CZ. Thus, from the point of view of tachocline modelling, it is not
unrealistic to regard the latitudinal differential rotation at $r=\rbcz$ as a
given boundary condition imposed at the top of the region of interest. This is
tantamount to assuming that differential rotation is driven by a highly
effective mechanism in the convective zone which is not seriously influenced by
the processes in the tachocline.

As the layer studied is thin, we also adopt a plane parallel representation
for  it, with constant density. (The effects of the stable density
stratification are only implicitly taken into account by its role in
determining the turbulent viscosity, eq.~(\ref{eq:nucr}).)

Thus, we regard the following model problem. 
Consider a plane parallel layer of incompressible fluid of density $\rho$,
where the viscosity $\nu$ and the magnetic diffusivity $\eta$ depend on $z$ 
only. At $z=0$ where $z$ is the vertical coordinate (corresponding to depth
in the solar application we have in mind) a periodic horizontal shearing flow
is imposed in the $y$ direction:
\[ v_{y0}= v_0 \cos(kx) \]
(so that $x$ will correspond to heliographic latitude, while $y$ to the
longitude). We assume a two-dimensional flow pattern ($\partial/\partial y=0$) 
and
$v_x=v_z=0$ (no ``meridional flow''). An oscillatory horizontal ``poloidal''
field is is present in the $x$ direction, given by
\< V_p=\frac 1{(4\pi\rho)^{1/2}}\pdv Az               \label{eq:Vpdef}  \>
(in Alfv\'enic units).
The ``toroidal'' (i.e. $y$) component $A$ of the vector potential obeys the 
corresponding component of the integral of the induction equation, which in our
case simplifies to a diffusion equation:
\< \pdv At=\pdv{}z\left(\eta\pdv Az\right)  \label{eq:A}
\>
The upper boundary condition is $A=A_0\cos(\omega t)$ at $z=0$, where $A_0$
fixes the poloidal field amplitude.

The evolution of the azimuthal components of the velocity and the magnetic
field is described by the corresponding components of the equations of
motion and induction, respectively. Introducing $v=v_y$ and using Alfv\'en
speed units also for the toroidal magnetic field
\<  b= B_y(4\pi\rho)^{-1/2} , \>
these can be written as
\< \pdv vt= V_p \cos(\omega t)\pdv bx +
   \pdv{}z\left(\nu\pdv vz\right)         \label{eq:veq} \>
\< \pdv bt= V_p \cos(\omega t)\pdv vx +   
   \pdv{}z\left(\eta\pdv bz\right)        \label{eq:beq} \>
where we have taken into account that, owing to the thinness of the tachocline,
the vertical derivatives dominate the diffusive terms.\footnote{The large-scale 
quasi-2D motions are assumed not to contribute to angular momentum transport
here, as the amplitude and sense of such a contribution is still highly
controversial, cf. \citeN{Spiegel+Zahn}, \citeN{Gough+McIntyre}.}
As the imposed poloidal field $V_p$ is independent of $x$, Fourier
transforming (\ref{eq:veq}) and (\ref{eq:beq}) in terms of $x$ yields the same
equations for the Fourier amplitudes $\hat v$ and $\hat b$, 
except that $\partial/\partial x$ is substituted by $ik$:
\< \pdv vt= ikb V_p \cos(\omega t) +
   \pdv{}z\left(\nu\pdv vz\right)         \label{eq:veq2} \>
\< \pdv bt= ikv V_p \cos(\omega t) +   
   \pdv{}z\left(\eta\pdv bz\right)        \label{eq:beq2} \>
(Hats are omitted to simplify notation.)
For a rough estimate, we write $\pi b/P$ for the l.h.s. of (\ref{eq:beq2}), then
substitute the resulting expression of $b$ into (\ref{eq:veq2}), take the real
part, and omit the factor $\cos^2(\omega t)$ in the first term on the r.h.s.:
\< \pdv vt= -k^2 V_p^2 P v +
   \pdv{}z\left(\nu\pdv vz\right)         \label{eq:veq3} \>
The equations to solve are thus (\ref{eq:Vpdef}), (\ref{eq:A}) and 
(\ref{eq:veq3}), with the turbulent diffusivities $\nu=\eta=\nu_m+\nucr$ given
by equation (\ref{eq:nucr}) with the identification $v_h=v$. $\nu_m$ is a
minimal diffusivity value (``molecular diffusivity'').

The simplified form of the first term in equation (\ref{eq:veq3}) will not
allow a correct reproduction of the periodic part of the time dependence. The
important point is, however, that this sink term, representing the reduction of
horizontal shear by Maxwell stresses, has the right amplitude and the correct
scaling with $V_p$, $P$, $v$, and $k$, so it may be expected to reproduce well
the cycle-averaged flow amplitude as a function of $z$, which is our main
interest here. Indeed, solving our equations for the case of constant
diffusivities $\nu$ and $\eta$, the results are in a remarkably good agreement
with the full 2D solutions in spherical geometry, presented in
\citeN{Dajka+Petrovay:fast1}. 

   \begin{figure}
   \centering
   \epsfig{width=1.3\linewidth,angle=90,figure=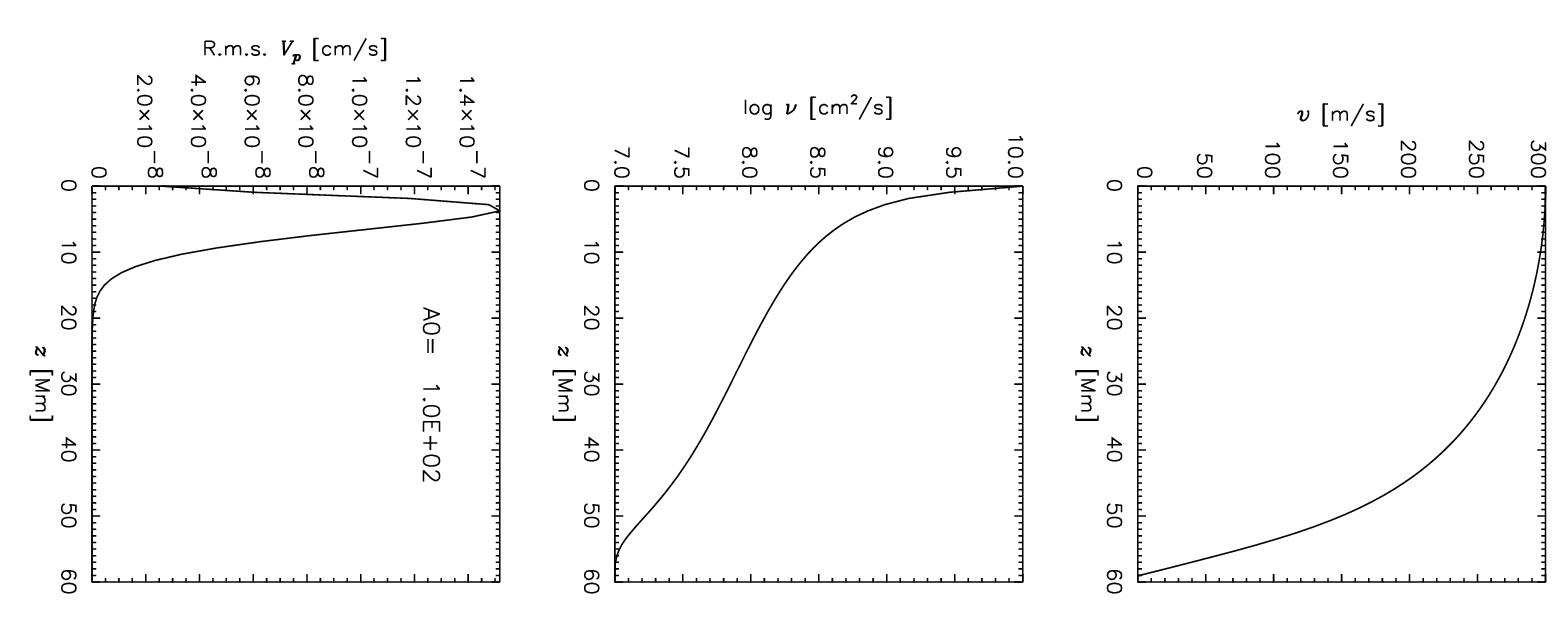}\hfill
   \epsfig{width=1.3\linewidth,angle=90,figure=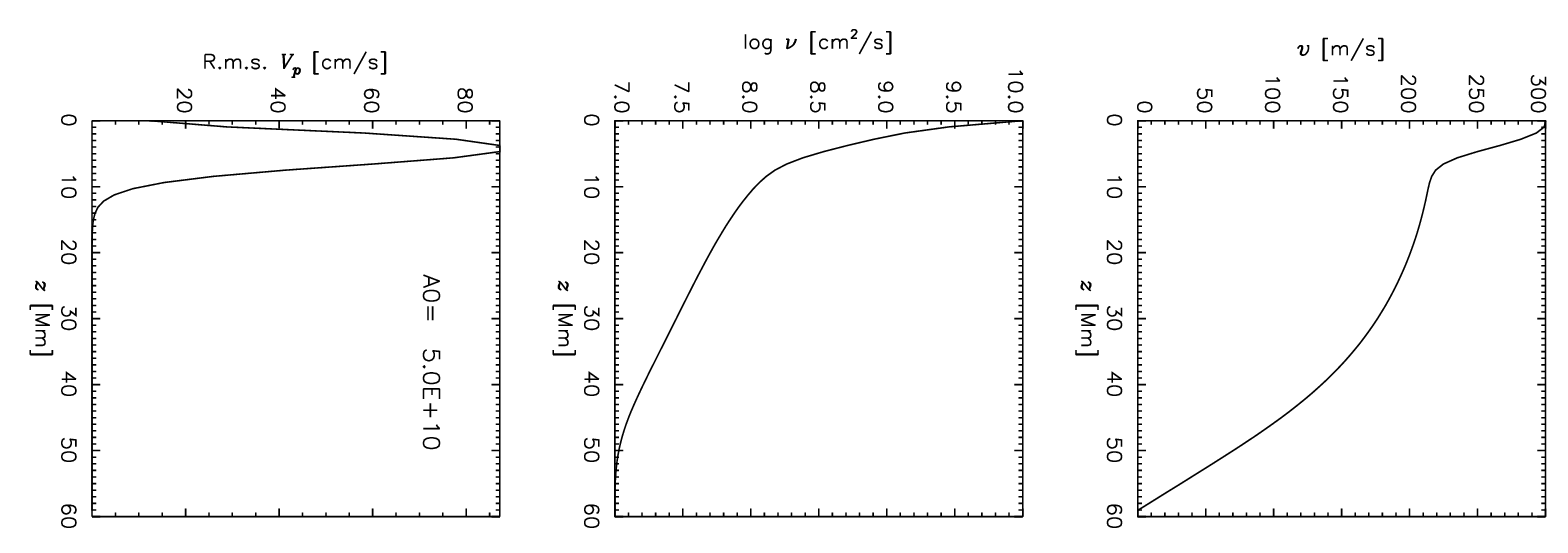}
      \caption{Horizontal differential rotation amplitude $v=v_h$, as defined 
      in eq. (\ref{eq:vhdef}) {\it (top row),} vertical turbulent diffusivity
      $\nu$ {\it (middle row),} and poloidal magnetic field in Alfvenic units
      {\it (bottom row)}, averaged over a solar half-cycle, as functions of
      depth below the convective zone, for a very low {\it (left-hand column)}
      and a medium {\it (right-hand column)} value of the field strength
      imposed at the top. Note that by coincidence, in the solar tachocline the
      field strength in gauss is roughly equal to the corresponding Alfven
      speed in cm/s.}
         \label{fig:results1}
   \end{figure}

   \begin{figure}
   \centering
   \epsfig{width=1.3\linewidth,angle=90,figure=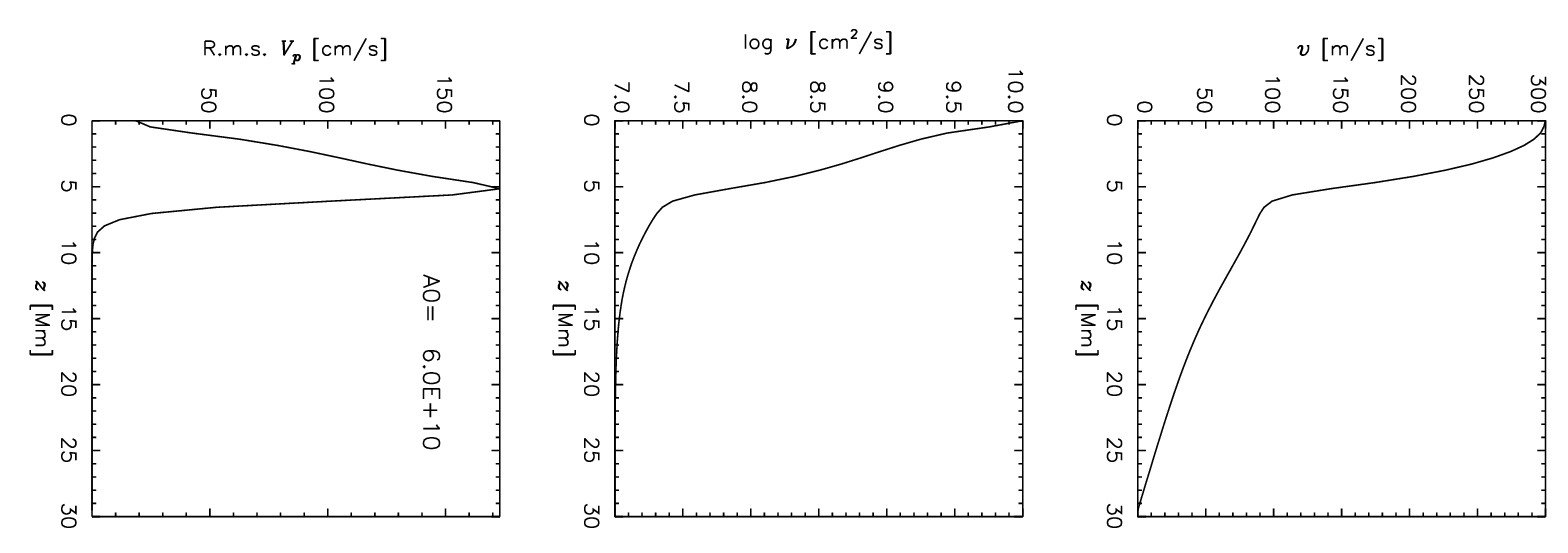}\hfill
   \epsfig{width=1.3\linewidth,angle=90,figure=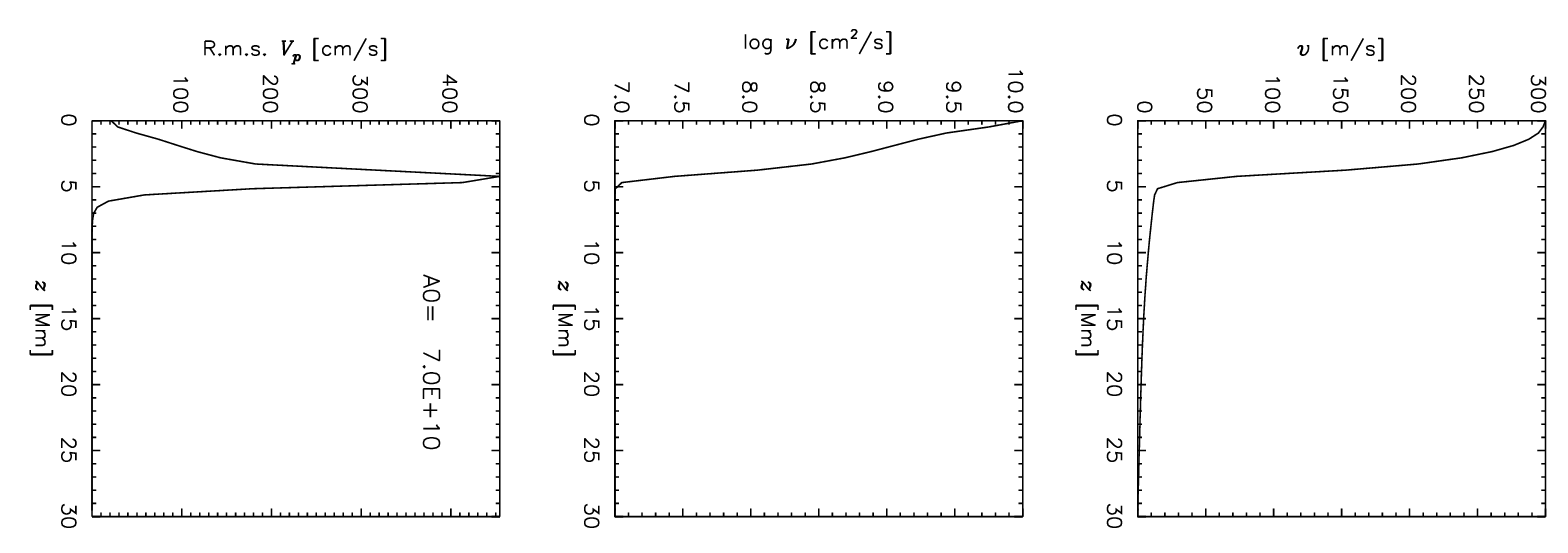}
      \caption{Same as Fig.~\ref{fig:results1} for two higher field strength values.
              }
         \label{fig:results2}
   \end{figure}

The equations were solved numerically by a finite difference scheme
second-order accurate in time. All quantities were set to zero at the lower
boundary, situated at $z_0=60$ or $30$ Mm below $\rbcz$, while the boundary
conditions applied at top ($z=0$) were $v=v_0=3\cdot 10^4\,$cm/s and
different prescribed values for $A_0$.  As equation~(\ref{eq:nucr}) is singular
at $z=0$, $\nu$ was set to a high  finite value $\numax$ here. For $\nu_m$ we
used the value $10^7\,$cm$^2/$s. This is much higher than the actual molecular
diffusivities in the tachocline, but using a realistic value would lead to
forbiddingly long integration times. Similarly, a too high value for $\numax$
would lead to very short timesteps, also increasing the computing time to
unaffordable values. Test runs with varying values of $\numax$ and $\nu_m$,
however, show that these choices do not significantly distort the results.

Starting from an arbitrary intial state, the system was allowed to evolve until
very nearly periodic behaviour sets in (in about $10^4$ years, depending on the
value of $\nu_m$), then average quatities for one 11-year half-cycle were
computed and plotted as functions of depth (Figs.~\ref{fig:results1} and
\ref{fig:results2}). 

\subsection{Results}

In the case with a very weak magnetic field (left-hand column in
Fig.~\ref{fig:results1}), it is straightforward to show that equation
(\ref{eq:veq3}) with $\nu$ given by (\ref{eq:nucr})  admits the analytic
solution
\<  v=v_0\left[ 1-\left(\frac z{z_0}\right)^2\right]^{1/4} ,
       \label{eq:analsol}  \>
confirmed by the numerical calculations. This essentially means that in this
case the shear penetrates as far down into the radiative interior as the
placement of the lower boundary condition allows. The weak magnetic field itself
only penetrates down to the skin depth given by equation (\ref{eq:skin}), as
expected. It is worth noting that the poloidal field amplitude shows a
non-monotonic behaviour with depth in all cases, reaching its maximum at some
finite $z$ value. This is due to the variable diffusivity: the horizontal field
lines tend to ``pile up'' where the diffusivity is significantly reduced.

From the right-hand column of Fig.~\ref{fig:results2} we can see that a
poloidal magnetic field of a few hundred gauss (peak strength 500 G) can
confine the tachocline to a thickness of barely 5 Mm. 

One might think that an intermediate field strength might lead to a somewhat
thicker tachocline. This is, however, not the case: an inspection of the full
series of results in our figures clearly shows that a weaker field simply
results in an ``aborted tachocline'', i.e. the horizontal shear is first
reduced by a factor depending on the field strength in a thin layer of a few
Mm, but below that layer, as the magnetic field is damped by the skin effect,
it follows the field-free solution (\ref{eq:analsol}), with a lower amplitude.

\section{Discussion: conditions of validity}

\subsection{Scales of the horizontal motions}
One important assumption in our model was that the velocity and time scales of
horizontal motions due to the shear instability are given by equations
(\ref{eq:vhdef}) and (\ref{eq:thdef}). This is indeed the general experience
with unstable shear flows, its physical background being an equilibrium 
between growth and turbulent dissipation of the dominant modes. The growth
timescale is 
\< \tau_g\sim l_h/\Delta v \label{eq:taug} \>
while the dissipation timescale is
\< \tau_d\sim l_h^2/\nu_h  \>
Equating these timescales and substituting $\nu_h\sim l_h v_h$ for the turbulent
diffusivity, we indeed find $v_h\sim\Delta v$.

But this line of argument is na\"{\i}ve for several reasons. First, in the
strongly stratified, quasi-2D flow in the tachocline the main mechanism of
dissipation is not the nonlinear transfer between 2D modes but the viscous
dissipation (due to overturning small-scale 3D turbulence) between neighbouring
slabs. Thus, the dissipation timescale is rather
\< \tau_d\sim l_c^2/\nu . \>
Nevertheless, substituting here the formula (\ref{eq:lc}), derived assuming the
validity of equations (\ref{eq:vhdef}) and (\ref{eq:thdef}), the result is again
$v_h\sim\Delta v$, confirming that our assumption was consistent.

Secondly, equation (\ref{eq:taug}) may not be a valid expression of the growth 
timescale if the instability is only slightly supercritical. Linear 2D 
hydrodynamical stability analysis indeed shows that the horizontal shear in the
tachocline is close to marginal stability
(\citeNP{Dziembowski+Kosovichev:difrot.stab}, 
\citeNP{Charbonneau+:tacho.HDstab}). The weakly nonlinear extension of the
analysis (\citeNP{Garaud:tacho.instab}) indicates that the evolution of
unstable modes is such that they modify the mean shear profile towards
increasing stability, thus ensuring marginal stability, low growth rates and
low velocity amplitudes. As, however, 3D effects and magnetic fields are known
to lead to strong instability with much weaker shear
(\citeNP{Gilman+Dikpati:unstable.tacho},
\citeNP{Dikpati+Gilman:shallow.alpha}), one may expect that this
near-criticality is a feature limited to 2D HD models only.

Somewhat surprisingly, the 3D thin shell simulations of \citeN{Miesch:tacho2}
also show rather low nonaxisymmetric horizontal flow amplitudes, at least one
order of magnitude below the shear amplitude. This is so despite the strongly
supercritical rotation profile chosen, which is not significantly altered by
the feedback of the nonaxisymmetric flow. One may speculate that this is the
result of the rather high numerical viscosities applied.

Further study of the flows driven by the horizontal shear instability is
clearly important. In addition to clarifying the issue of characteristic scales
of this flow, such studies may shed more light on the question of the
importance and sense of any direct angular momentum transport by these motions.
At any rate, a lower velocity amplitude or a longer timescale for the
horizontal flows would result in lower critical field strengths for the
secondary shear instability, resulting in an even thinner tachocline confined
by even weaker fields in the framework of the present model.

\subsection{Other sources of turbulence}
Another important constraint in the present model was the assumption that the
secondary shear instability is the only source of turbulence in the tachocline.
While a number of other instabilities in rotating radiative stellar interiors
are known (cf. \citeNP{Tassoul:book}, \citeNP{Zahn:long},
\citeNP{Spruit:interior.instab}), they are mostly acting throughout the
radiative interior where constraints such as chemical mixing in stellar models
impose rather strict limits on turbulence. Thus, in the tachocline we need
``extra'' sources of turbulence, for which shear instabilities are the most
plausible candidate.

More important than rotational instabilities is the possibility of nonadiabatic
overshooting convection. The recent solution of the Reynolds moment equations
for the overshoot layer by \citeN{Marik+Petrovay:isotropic} indicates that the
total extent of nonadiabatic overshoot is negligible. On the other hand, 
\citeN{Xiong+Deng:ushoot} find a very significant nonadiabatic overshoot. As
both models have their own shortcomings, thgis issue is not quite settled yet.
At any rate, if there is a significant nonadiabatic overshoot, the secondary
shear instability would only set in below the overshoot layer, where the
turbulent diffusivity is reduced below the critical value; in this layer, our
model could still be considered valid. Thus, the main effect of any
nonadiabatic overshoot would be to ``shift'' the resulting tachocline to
somewhat lower depths (unless a stronger dynamo field reduces the horizontal
shear already in the overshoot region).

\section{Conclusion}

This paper presented the first consistent model for the turbulent tacho\-cline,
with the turbulent diffusivity computed within the model instead of being
specified arbitrarily. For the origin of the 3D turbulence a new mechanism was
proposed, the secondary vertical shear instability of 2D motions taking place
in thin horizontal slabs. 

Independent evidence for the mechanism proposed here is hard to find, as the
conditions in the tachocline are extreme by terrestrial standards, and direct
numerical simulations of stratified shear flows are currently limited to much
lower values of the Richardson and Reynolds numbers. Nevertheless, the presence
of a quasi-2D, layered structure, and a strongly anisotropical turbulent
diffusivity tensor have been confirmed in numerical simulations by
\citeN{Metais+Herring}, \citeN{Kaltenbach+}  and 
\citeN{Jacobitz+Sarkar:horiz.shear}. Indications for the presence of a
secondary shear instability were also found in the recent simulations of
\citeN{Werne+Fritts}.

A formula for the diffusivity due to the turbulence generated by the secondary
shear, equation (\ref{eq:nucr}) was derived and applied in a simple 1D model of
the tachocline, taking into account the Maxwell stresses due to the dynamo
field. As our analysis does not consider the effects of spherical geometry,
rotation, non-adiabatic convective overshoot, meridional circulation, or the
effect of magnetic fields on  stability, it should only be regarded as a first
step towards a more comprehensive theoretical analysis of the problem of
turbulence in the solar tachocline.

A thin tachocline in our models can only be produced if the oscillatory
dynamo magnetic field exceeds a critical limit. At this limit, the thickness of
the tachocline is $\sim 5\,$Mm and the mean value of the field strength in the
tachocline is 200 G. The total magnetic flux in the tachocline in this case
agrees well with the total poloidal flux crossing the equatorial plane in the
poloidal flux transport model of \citeN{Petrovay+Szakaly:2d.pol}. 

On the other hand, the tachocline resulting from our model is uncomfortably
thin when compared to helioseismic constraints. Existing seismic calibrations
of tachocline thickness (\citeNP{Basu+Antia:tachovar}) indicate a significantly
thicker shear layer. These calibrations are based on the use of simple 
predefined fitting profiles, characterized by a single length scale. In
contrast, our model indicates that for slightly subcritical magnetic field
strengths the profile is more complicated, the fast linear cutoff in $v$ within
the skin layer being replaced by a shallow but deep profile of type
(\ref{eq:analsol}) below. It remains to be seen, whether such a two-tiered
$v$-profile can yield an equal or better fit to helioseismic data than the more
conventional profiles.

The thin tachocline resulting from the present model is also hard to reconcile
with the gradual depletion of lithium in the atmospheres of Sun-like stars
during their lifetimes. Lithium is destroyed by nuclear reactions in layers
below $z\sim 40\,$Mm only, so a mixing characterized by a diffusivity of at
least $10^3\,$cm$^2/$s must be present as far down as that depth. While our
prescription $\nu_m=10^7\,$cm$^2/$s does not allow a firm statement on this
issue, the very sharp cutoff of the $\nu$-curve in the right-hand column of
Fig.~\ref{fig:results2} does not seem to indicate that any significant level of
turbulence could be maintained at such great depths. The two-tiered character
of our profiles might admit some ``leakage'' of turbulence into much deeper
layers, but this would imply some rather implausible fine-tuning of the
dynamo field strength.

One obvious shortcoming of the present models is their simplified treatment of
the time development and of the geometry. The development of more realistic, 
axially symmetric spherical models employing the viscosity formula
(\ref{eq:nucr}) is in progress.

\acknowledgements 
I am grateful for the hospitality of the Center for Turbulence Research at 
NASA Ames Research Center during part of this work. This paper owes much to 
discussions with A. Kosovichev, N. Mansour, M. Miesch and A. Wray. 
This work was supported by the OTKA under grants no.\ T032462 and T034998.


\begin{ao}
\\
K. Petrovay\\
E\"otv\"os University, Dept.~of Astronomy\\
Budapest, Pf.~32, H-1518 Hungary\\
E-mail: kris@astro.elte.hu
\end{ao}                   

\end{article}
\end{document}